\documentclass[prl,twocolumn]{revtex4}

\def\be{\begin{equation}}
\def\ee{\end{equation}}
\def\bea{\begin{eqnarray}}
\def\eea{\end{eqnarray}}

\def\npb#1#2#3{{\it Nucl.\ Phys.} {\bf B#1} (#2) #3}
\def\plb#1#2#3{{\it Phys.\ Lett.} {\bf B#1} (#2) #3}
\def\prl#1#2#3{{\it Phys.\ Rev.\ Lett.} {\bf #1} (#2) #3}
\def\prd#1#2#3{{\it Phys.\ Rev.} {\bf D#1} (#2) #3}

\def\apj#1#2#3{{\it Ap.\ J.} {\bf #1} (#2) #3}

\begin{document}
\renewcommand{\topfraction}{0.99}
\renewcommand{\bottomfraction}{0.99}

\title{Astrophysical violations of the Kerr bound \\ 
as a possible signature of string theory}
\author{Eric G. Gimon and Petr Ho\v{r}ava}
\affiliation{
Berkeley Center for Theoretical Physics and Department of Physics,\\
University of California, Berkeley, California 94720-7300\\ 
and\\
Theoretical Physics, Lawrence Berkeley National Laboratory,\\
Berkeley, California 94720-8162}
\date{\today}
\begin{abstract}
In 4D general relativity, the angular momentum of a black hole is limited by 
the Kerr bound.  We suggest that in string theory, this bound can be 
breached and compact black-hole-like objects can spin faster.  Near such 
``superspinars,'' the efficiency of energy transfer from the accreting matter 
to radiation can reach 100\%, compared to the maximum efficiency of 42\%\ of 
the extremal Kerr (or 6\%\ of the Schwarzschild) black hole.  Finding such 
superspinning objects as active galactic nuclei, GBHCs, or sources of gamma 
ray bursts, could be viewed as experimental support for string theory.
\end{abstract}


\maketitle
\vskip2pc]
String theory is a popular candidate for quantum theory of gravity, but robust 
experimental tests still seem out of reach.  Two areas where string 
theory could make contact with the real world are usually suggested: 
high-energy particle physics, and cosmology.  However, cosmological 
questions require the control of time-dependent backgrounds and spacelike 
singularities, an area where string theory is still rather weak.  Similarly, 
robust predictions for accelerator physics are tied to the 
problem of the choice of a vacuum solution, also relatively poorly understood 
despite recent progress.  

This paper advocates the point of view that string theory should be 
applied to areas where it has already demonstrated its strengths.  In 
particular, it has proven exceptionally good at resolving spacetime geometries 
with various timelike singularities.  (The theoretical control over such 
solutions increases with the increasing degree of spacetime supersymmetry 
(SUSY).)  Such singularities, inconsistent in general relativity (GR), then 
represent new classes of legitimate compact objects in the string-theory 
completion of GR.  We suggest that such objects may be relevant for the 
observational astrophysics of compact objects \cite{krolik,compbook,putten}.  
Relativistic astrophysics is thus another, hitherto underrepresented, area of 
physics where signatures of string theory should be sought.  

Clearly, not all timelike singularities of GR are consistently resolved in 
string theory.  (An example believed to stay pathological in string theory 
is the negative-mass Schwarzschild metric.)  In this paper, we take advantage 
of the recent 
progress in understanding objects with angular momentum in string theory.  
Specifically, we concentrate on the possibility of violating the 
Kerr bound on the angular momentum carried by compact objects.  
The existence of such ``superspinars'' would have significant observational 
consequences, in particular for AGNs, GBHCs and GRBs.  The basic question 
becomes an experimental one: {\it Do we observe candidate compact objects that 
violate the Kerr bound?  If so, they can find a natural interpretation in 
string theory.}

In four spacetime dimensions, spinning black holes with specific angular 
momentum $a=J/M$ are described by the famous Kerr solution of GR, given in 
the Boyer-Lindquist (BL) coordinates (with $G_N=c=1$) by
\bea
\label{kerrs}
ds^2=&&-\left(1-\frac{2Mr}{\Sigma}\right)dt^2-\frac{4aMr\sin^2\theta}{\Sigma}
dt\,d\phi+\frac{\Sigma}{\Delta}dr^2\nonumber\\
{}+&&\Sigma\,d\theta^2+\left(r^2+a^2+\frac{2Mra^2\sin^2\theta}{\Sigma}\right)
\sin^2\theta\,d\phi^2,
\eea
with $\Sigma=r^2+a^2\cos^2\theta$ and $\Delta=r^2-2Mr+a^2$.  
In the case of black holes carrying an electric charge $Q$, the relevant 
solution would be the Kerr-Newman black hole.  The absence of naked 
singularities leads to the Kerr bound,
\be
\label{kerrb}
a^2+Q^2\leq M^2.
\ee
In GR, some form of cosmic censorship is usually assumed, and the 
over-rotating solutions are discarded as unphysical.  

The Kerr-Newman family can be embedded, via minimal $N=2$ supergravity, into 
string theory.  In the supersymmetric setting, one encounters another 
important bound: the BPS bound,
\be
\label{bpsb}
Q^2\leq M^2.
\ee
There is an interesting clash between the two notions of extremality: The 
SUSY notion implied by (\ref{bpsb}) and the more restrictive GR one suggested 
by (\ref{kerrb}).  The status of the BPS bound is much stronger than that of 
the Kerr bound: It is an {\it exact\/} bound, implied in a supersymmetric 
vacuum by the kinematics of the supersymmetry algebra.  On the other hand, the 
Kerr bound is a consequence of the detailed prejudice about the regions of 
very strong curvature (for example, if one assumes the exact validity of 
classical 
GR); it should thus be viewed as an approximate bound and expected to receive 
substantial corrections in string theory, possibly enlarging the space of 
asymptotically observable parameters (such as $a$) that correspond to 
legitimate compact objects.  In this sense, the cosmic censorship conjecture 
would be invalid in its most naive GR form: Some ``naked singularities'' 
of GR would be legitimate not because they hide behind horizons, but because 
they are resolved due to high-energy effects of the deeper theory.  It is 
indeed important to apply to astrophysical objects lessons learned in 
effective field theory: Observations at a given energy scales (or spacetime 
curvature) should not require detailed knowledge of the physics at a much 
higher energy scale (or curvature).  Imposing standard cosmic censorship of GR 
on astrophysical objects violates this ``decoupling principle,'' 
by extrapolating GR into the high-curvature regime.  

As our first example where a breach of the Kerr bound is achieved in 
string theory in a controlled setting, consider a class of SUSY solutions in 
$4+1$ dimensions known as the ``BMPV black holes'' \cite{bmpv}.  They are 
solutions of minimal supergravity, with self-dual angular momentum $J$, 
mass $M$ and electric charge $Q$.  The BPS bound requires $|Q|=M$, but puts 
no restriction on $J$.  The BMPV solutions satisfying the BPS bound 
have a horizon area $A=\sqrt{Q^3-J^2}$.  The Kerr bound analog thus requires 
\be
\label{bmpvkerr}
J^2\leq Q^3.
\ee
The SUSY BPMV solutions are extremal black holes for $J\leq Q^{3/2}$, and 
naked singularities for $J> Q^{3/2}$.  In fact, the situation is even worse: 
The naked singularity is surrounded by a compact region of closed timelike 
curves, making it a naked time machine.  It appears puzzling that string 
theory contains a perfectly supersymmetric solution with such apparently 
pathological causal features.  

A stringy resolution of this paradox has been presented in \cite{gh}.  The 
pathological core of the solution is excised by a spherical domain wall 
made out of microscopic constituents ({\it i.e.}, strings and D-branes) that 
carry the same total charge and mass as the naked BMPV solution.  The outside 
geometry stays intact, but the pathological inside has been replaced by a 
causal portion of the G\"odel universe solution \cite{bghv}.  In this way, 
the BMPV solution and the G\"odel solution solve each other's causality 
problems!  The dynamics of the domain wall implies that consistent solutions 
now satisfy 
\be
J^2\leq (Q+R)^3,
\ee
where $R$ is the radius of the domain wall.  In this way, the Kerr bound 
(\ref{bmpvkerr}) has been relaxed due to a stringy effect, allowing a larger 
class of compact objects in string theory compared to classical GR.  Note 
that there is now no bound on $J$, but for large enough $J$ the domain wall 
becomes so large that the object is not inside its Schwarzschild radius, 
and is no longer sufficiently ``compact''.  (Clearly, objects larger than 
their Schwarzschild radius are not subject to the Kerr bound.  The violation 
of the Kerr bound is of interest only for objects sufficiently compact, and 
we restrict our attention to those.)  Note also that even for $J$'s that 
violate the original bound (\ref{bmpvkerr}), the resolved solution is 
described at long distances by the BMPV metric, valid all the way to the 
domain wall.   There, supergravity is finally modified, just before one 
extrapolates the GR solution to the pathological region.  

Once the Kerr bound has been breached in the supersymmetric setting and in 
$4+1$ spacetime dimensions, one should expect this phenomenon to be generic 
in string theory; it should extend to $3+1$ dimensions and to vacua with 
supersymmetry either spontaneously broken or absent altogether.  Various such 
solutions are indeed known.  The solutions of heterotic string theory found 
in \cite{ganda} look from large distances $3+1$ dimensional, and as if sourced 
by a naked over-rotating singularity.  However, their core is resolved 
by Kaluza-Klein modes of two extra dimensions compactified on a torus:   
Instead of a naked singularity, the core contains a 
periodic array of black holes along the compact dimensions.  Thus, an object 
violating the Kerr bound of GR in $3+1$ dimensions becomes legitimate in 
string theory.  Another example comes from reducing black ring solutions such 
as \cite{elvang} with residual angular momentum.  These reductions typically 
have a multi-center structure and extra $U(1)$ fields 
which drive angular momentum up via dipole moments.  Outside of SUSY, there 
exist extremal Kaluza-Klein black holes \cite{rasheed,larsen} with a slow 
rotation phase and a fast rotation phase separated by a singular 
configuration.  This list is far from exhaustive, but the known solutions 
have a common thread: A geometry described by a forbidden GR solution is 
resolved by high-energy GR modifications at the core, typically via brane 
expansion and the formation of an extended bound state with dipole moments.   

Having established the possibility of breaching the Kerr bound for compact 
objects in string theory, we now return to our Universe of $3+1$ spacetime 
dimensions.  Realistic astrophysical black holes are uncharged, and thus 
described -- in the long-distance regime where GR holds -- by the Kerr 
solution (\ref{kerrs}).  At this stage, string theory does not allow enough 
control over solutions far from SUSY; hence we cannot present a detailed 
mechanism for resolving the over-rotating Kerr singularity with the level of 
rigor comparable to the examples above.  Given the lessons learned from those 
examples, however, we will {\it assume\/} that the Kerr black hole in string 
theory behaves similarly, {\it i.e.},  we assume 
that a resolution of the Kerr singularity exists for angular momenta 
violating the Kerr bound and with the metric outside of the central stringy 
core given by (\ref{kerrs}) in the over-rotating regime $a^2>M^2$.  

In the vicinity of the singularity, we expect large stringy modifications.  
Just like the over-rotating BMPV solution, the over-rotating Kerr would be 
a naked time machine, with closed timelike curves in the region $r<0$. 
If this solution is to be consistent, this pathological region must be excised 
or otherwise modified by stringy sources.  This form of chronology protection 
might be generic in string theory \cite{bghv,herdeiro}.  Thus, we 
assume that the region below some $r=\epsilon>0$ has been so modified.  

How much do we need to know about the detailed resolution mechanism at the 
core of the solution?  This question is related to the ``decoupling 
principle'' mentioned above.  Given the current absence of a precise model, 
we can only study questions that do not depend on the details of the 
resolution in the high-curvature region at the core.  Luckily, some of the 
most interesting astrophysical effects take place at distances sufficiently 
far from the core, where GR is valid and the details of the core dynamics are 
largely irrelevant.  This is so because of a fortunate feature of the Kerr 
metric in $3+1$ dimensions:  As we approach $a=M$ from below (by dialing the 
value of $a$, not by a physical process), the size of 
the horizon depends on $a$ in a manner reminiscent of the order parameter 
during a first-order phase transition; it approaches a nonzero limiting 
value $\sim M$, instead of shrinking to zero area and infinite curvature.  
This creates a useful separation of scales for compact objects with spins not 
too far above the Kerr bound: The astrophysically interesting regime will be 
at $r\sim M$, far from the stringy core for $M$ of astrophysical interest.  

If such ``Kerr superspinars'' existed as compact objects in our Universe, 
how would one observe them?  Just as we search for black hole candidates, 
we should search for superspinars in a number of astrophysically relevant 
situations: long-lived ones as candidates for active galactic nuclei (AGN) 
\cite{krolik} or galactic GBHCs \cite{fabian}, and those that develop an 
instability as possible mechanisms for GRBs \cite{putten}.  

In the case of AGNs, the main reason supporting the black-hole paradigm 
\cite{compbook,krolik,rees} is their exceptionally high luminosity, which 
suggests that they are powered by an accretion process of high efficiency.  
We claim that superspinars would likely be among the most luminous objects of 
comparable mass.  The simplest model of energy production by compact objects 
assumes a thin accretion disk along the equatorial plane.  Accreting matter 
moves along direct circular stable orbits, losing angular momentum slowly due 
to viscosity.  In the process, its energy is radiated to infinity, until 
it reaches the innermost stable circular orbit (ISCO).  Then it plunges into  
the black hole and the remainder of the rest mass is lost.  The efficiency of 
the process is thus measured by the rest mass at the ISCO 
\cite{bardeen,compbook}.  Any realistic situation is likely much more complex, 
but for our purposes it will suffice to adopt this simple picture.  

For the Schwarzschild black hole, the ISCO is at $r=6M$, reaching closer to 
the horizon with increasing $a$ of the Kerr solution, all the way to $r=M$ at 
$a=M$ \cite{bardeen,compbook}.  (This is a well-known artefact of the 
breakdown of BL coordinates at $a=M$; the ISCO at extremality is still a 
nonzero proper distance above the horizon.)  The efficiency of accretion rises 
from $\sim 6\%$ at $a=0$ to $1-1/\sqrt{3}\sim 42\%$ at extremality.  

For superspinars, the equatorial orbits are governed by the effective 
potential 
\bea
\nonumber
V(r)&=&\frac{L^2}{2r^2}-\frac{M}{r}+\frac{1-E^2}{2}\left(1+\frac{a^2}{r^2}
\right)\nonumber\\ 
&&\qquad{}
-\frac{M}{r^3}(L-aE)^2,
\eea
with $L$ the specific angular momentum and $E$ the energy at infinity per unit 
mass of the probe.  The ISCO is where the two roots of $V'(r)$ coincide while 
$V(r)=0$.  It is particularly interesting to look for the value of $a/M$ at 
which the efficiency of accretion reaches $100\%$.  This happens when the ISCO 
has $E=0$, at
\be
\label{hundred}
a/M=\sqrt{32/27}\approx 1.0886\ ,
\ee
{\it i.e.}, for over-rotation by less than $9\%$.  At that point, the ISCO has 
been dragged even deeper into the solution, to $r=2M/3$.  (Stangely, the same 
ratio 32/27 makes an appearance in $4+1$ dimensions, as the ratio of the 
maximum angular momentum of the neutral black hole and the minimal angular 
momentum of the black ring.) This is the minimal value of $r$ that the ISCO 
can reach as $a$ varies.  For a supermassive superspinar, the ISCO is in the 
region of relatively weak curvature at large proper distance from the 
(resolved) singularity, hence insensitive to the details of the dynamics of 
the core in string theory.  A particle at the direct ISCO for $a/M$ given by 
(\ref{hundred}) carries negative specific angular momentum $L=-2M/\sqrt{27}$ 
and its accretion will lower the value of $a$.  Thus, accretion of matter from 
ISCO will generally push $a/M$ of superpinars from higher to lower values.  In 
any case, due to the lowering of the ISCO and consequently the very high 
efficiency of the accretion process, superspinars are likely to be very 
luminous during their active phase.

The available data strongly suggest that most AGNs will carry significantly 
high values of $a/M$, perhaps requiring on average as much as $15\%$ 
efficiency \cite{elvis}.  Many objects spin significantly faster 
\cite{breynolds,fabian}, near the Kerr bound (or its refinement due to Thorne 
\cite{thorne}).  Among AGNs, one of the most famous examples is the Seyfert~I 
galaxy MCG-6-30-15 \cite{iwasawa}; some GBHCs are also rapidly rotating 
(see, {\it e.g.},\cite{fabian}), for example XTE J1650-500.  

A useful signature of the specific angular momentum of compact objects comes 
from their X-ray spectroscopy, in particular, the shape of the $\sim 6.4$ keV  
fluorescent Fe K$\alpha$ emission lines associated with the innermost 
regions of the accretion disk \cite{laor,nowak}.  In a number of notable 
examples, very broad red wings of these lines have been observed, strongly 
suggesting high values of $a$.  While such findings are typically analyzed 
under the strict assumption that the Kerr bound holds, it would be interesting 
to see if some of the observations are compatible with the emission from the 
accretion disk near the ISCO of a slightly over-rotating superspinar.  Future 
missions, in particular {\it Constellation-X\/} and {\it LISA}, will provide 
crucial data leading to the measurement of angular momenta of compact objects 
with larger accuracy, possibly testing whether any of them breach the Kerr 
bound.  

We conclude this paper with miscellaneous comments: 

$\bullet$ Kerr superspinars do not have event horizons, but when they spin 
only slightly above the Kerr bound they exhibit such a deep gravitational well 
that any escape requires essentially cosmological time scales.  A less 
teleological concept replacing the event horizon may be needed to describe 
this properly.  It turns out that the Kerr superspinar has a surface 
$\cal S$ at $r\sim M$  such that the expansion of null geodesics orthogonal 
to $\cal S$ and pointing outwards (to larger $r$) at fixed $t$ is zero.  Thus, 
the time-like surface $\cal S$ can play the role of a holographic screen -- an 
important concept in quantum gravity and string theory -- even though it does 
not satisfy the definition \cite{ashtekar} of a dynamical horizon.  

$\bullet$ Despite the absence of the event horizon, the Kerr superspinar 
maintains an ergoregion $\cal E$, with boundary at 
$r=M\pm\sqrt{M^2-a^2\cos^2\theta}$. For $a>M$ this only has solutions for 
$\theta$ smaller than some $\theta_c<\pi$: The ergoregion now fills a torus, 
with openings along the axis of rotation. 
Interestingly, this phenomenon could facilitate the formation of relativistic 
jets. We envision a ``tea-kettle effect'': Particles falling in from ISCO 
are trapped by the gravitational well, producing high pressure in the central 
region, with the natural escape route along the rotation axis where particles 
do not have to overcome the frame dragging of the ergoregion.  This could be 
further enhanced by the spin-orbit interaction between the accreting matter 
and the superspinar \cite{putten}, and by electromagnetic effects in 
magnetized accretion disks, such as those involved in the Blandford-Znajek 
process.  

$\bullet$ A next step beyond the scope of this paper is the study of possible 
instabilities of superspinars, which can be two-fold: Those of the stringy 
core, requiring a detailed stringy model; and the universal low-energy 
instabilities visible in GR approximation, such as the ergoregion 
instability \cite{ergoinst}.  

$\bullet$ A string-theory model of superpinars might involve a spherical 
domain wall similar to that of the $4+1$ example above.  However, an even 
simpler option suggests itself:  Could F-strings or D-strings serve as 
candidate superspinars?  Recall that string theory contains another 
fundamental bound: the Regge bound relating the mass and angular momentum 
of F-strings.  Intriguingly, only in $3+1$ dimensions is the Regge bound 
qualitatively of the same form as the Kerr bound.  

$\bullet$ In classical GR, other solutions representing naked singularities 
with angular momentum are known, such as \cite{tosa}.  We have confined 
our attention only to Kerr superspinars, but other possibilities should also 
be studied.  

$\bullet$ Another question is that of the possible origin of superspinars.  
Can one overspin an existing Kerr black hole by an adiabatic physical 
process?   Preliminary analysis 
suggests that the answer is probably negative, and that superspinars and black 
holes should be viewed as two distinct phases of compact objects separated by 
a barrier at $a=M$.  This suggests that superspinars could not be created by 
collapse of ordinary matter in galaxies.  If so, superspinars in AGNs could 
still be primordial remnants of the high-energy phase of early cosmology when 
stringy effects were important.  This should be studied together with the 
time-scale problem \cite{krolik} in models of galaxy formation, which suggests 
the existence of rather massive seed nuclei in galaxies at surprisingly early 
times. 

We hope that the ideas presented in this paper will trigger further 
investigation in two independent directions: (1) the theoretical search for 
detailed models of superspinars in string theory or in phenomenological 
models; (2) an experimental search for observational signatures of 
superspinars in our Universe.  Finding compact objects that rotate faster than 
the Kerr bound would be a strong signature of strong-gravity modifications 
of GR.  String theory would then represent an arena in which such results 
could be naturally interpreted.  

We wish to thank T. Damour, C. Done, G. Horowitz, and E. Witten 
for useful discussions.  This work was supported by NSF Grants PHY-0244900 and 
PHY-0555662, DOE Grant DE-AC03-76SF00098, and the Berkeley Center for 
Theoretical Physics.  An early version of our arguments was presented by 
P.H. at {\it Strings 2004\/} in Paris.  


\end{document}